      [1999/12/01 v1.4c Il Nuovo Cimento]
\documentclass{cimento}

\newcommand{\be}{\begin{equation}}
\newcommand{\ee}{\end{equation}}
\newcommand{\bea}{\begin{eqnarray}}
\newcommand{\eea}{\end{eqnarray}}
\title{Low energy processes to distinguish among seesaw models}
\author{C.~Biggio}
\instlist{\inst Max-Planck-Institut f\"ur Physik, 80805 M\"unchen, Germany}
\PACSes{\PACSit{13.15.+g 14.60.Pq 14.60.St}{}}
\begin{document}

\maketitle

\begin{abstract}
We consider the three basic seesaw scenarios (with fermionic singlets,
scalar triplets or fermionic triplets) and discuss their
phenomenology, aside from neutrino masses. We use the effective field
theory approach and compare the dimension-six operators characteristic
of these models. We discuss the possibility of having large
dimension-six operators and small dimension-five (small neutrino
masses) without any fine-tuning, if the lepton number is violated at a
low energy scale. Finally, we discuss some peculiarities of the
phenomenology of the fermionic triplet seesaw model.
\end{abstract}

Neutrinos are exactly massless in the Standard Model (SM). However, in
the past decades, oscillation experiments have proved that neutrinos
are massive and nowadays we know they are lighter than few eV, thanks
to $^3H$ decay experiments, $0\nu\beta\beta$ decay experiments and,
more recently, to cosmology.

New physics (NP) is therefore required to explain neutrino masses and
a natural explanation of their smallness, with respect to charged
leptons ones, can be achieved in the context of seesaw models, which
introduce NP at an energy scale higher than the electroweak (EW)
one. A good way of considering the effects of this new high energy
physics is using the effective field theory approach. While with SM
fields many dimension-six ($d6$) and higher order operators can be
built, the $d5$ is unique and it is given by 
$\left[\frac{1}{2}\, c_{\alpha \beta}^{d5} \,
\left( \overline{\ell_L^c}_{\alpha} \tilde \phi^* \right) \left(
\tilde \phi^\dagger \, {\ell_L}_{ \beta} \right) + {\rm h.c}\right]$,
where $\ell_L$ are the lepton doublets and $\tilde \phi =
i \tau_{2} \phi^*$, with $\phi$ the Higgs field. Here $c_{\alpha
\beta}^{d5}$ is  a model-dependent coefficient matrix inversely 
proportional to the NP scale $M$. After EW symmetry breaking this
operator generates Majorana masses for neutrinos which, for large
enough $M$, are small for ``natural'' values of the
Yukawa couplings ($Y\sim O(10^{-6}-1)$).

The unique $d5$ operator can be generated by different types of NP. At
tree-level, there are only three possibilities: from heavy fermionic
singlets (type-I seesaw), scalar triplets (type-II seesaw) or
fermionic triplets (type-III seesaw). It is clear that from neutrino
masses it is impossible to understand which is the NP which lays
behind, so that, in order to distinguish among different models, other
associated effects have to be observed. $d6$ operators, which produce
them, have been computed in refs.~\cite{BGJ,Abada:2007ux},
respectively for the type-I model and types II and III, and are shown
in table~\ref{tableops}, together with the corresponding coefficients
(and the coefficients of the $d5$ operators). We immediately observe
that the $d6$ operators differ in the three models, so that in
principle, observing the physical effects they produce, it would be
possible to understand which is that one responsible for neutrino
masses if among these three.
\begin{table}
\caption{ Coefficients of the $d5$ operator, $c^{d5}$, and  $d6$ operators 
and their coefficients, $c^{d6}$, in the three basic seesaw theories.}
\begin{tabular}{cccc}
\hline
Model & $c^{d5}$ & $c^{d6}_{i}$ & $\mathcal{O}^{d6}_{i}$ \\
\hline
Fermionic Singlet & $Y_{N}^{T}\frac{1}{M_{N}}Y_{N}$ & $\left(Y_{N}^{\dagger}\frac{1}{M_{N}^\dagger}\frac{1}{ M_N}Y_{N}\right)_{\alpha\beta}$ & $\left(\overline{\ell_{L\alpha}}\widetilde{\phi}\right)i\partial\!\!\!/\left(\widetilde{\phi}^{\dagger}\ell_{L\beta}\right)$ \\
\hline
 & & $\frac{1}{ M_\Delta^{2}}Y_{\Delta \alpha\beta}Y_{\Delta \gamma\delta}^{\dagger}$ &$ \left(\overline{\widetilde{\ell_{L\alpha}}}\overrightarrow{\tau}\ell_{L\beta}\right)\left(\overline{\ell_{L\gamma}}\overrightarrow{\tau}\widetilde{\ell_{L\delta}}\right)$  \\[0.2cm]
Scalar Triplet & $4Y_{\Delta}\frac{\mu_{\Delta}}{M_{\Delta}^{2}}$ & $\frac{|\mu_{\Delta}|^{2}}{M_{\Delta}^{4}}$ & $\left(\phi^{\dagger}\overrightarrow{\tau}\widetilde{\phi}\right)\left(\overleftarrow{D_{\mu}}\overrightarrow{D^{\mu}}\right)\left(\widetilde{\phi}^{\dagger}\overrightarrow{\tau}\phi\right) $ \\[0.2cm]
  & & $-2\left(\lambda_{3}+\lambda_{5}\right)\frac{|\mu_\Delta|^{2}}{M_\Delta^{4}}$ & $\left(\phi^{\dagger}\phi\right)^{3}$   \\
\hline
Fermionic Triplet & $Y_{\Sigma}^{T}\frac{1}{M_{\Sigma}}Y_{\Sigma}$ & $\left(Y_{\Sigma}^{\dagger}\frac{1}{M_{\Sigma}^\dagger}\frac{1}{ M_\Sigma}Y_{\Sigma}\right)_{\alpha\beta}$ & $\left(\overline{\ell_{L\alpha}}\overrightarrow{\tau}\widetilde{\phi}\right)iD\!\!\!\!/\left(\widetilde{\phi}^{\dagger}\overrightarrow{\tau}\ell_{L\beta}\right)$  \\
\hline
\end{tabular}
\label{tableops}
\end{table}

Before starting the analysis of the phenomenology associated to $d6$
operators, we discuss here about the possibility of observing it. If
we consider $\mathcal{O}(1)$ Yukawa couplings and we look at the
fermionic seesaws, we see that $c^{d6}\sim(c^{d5})^2$. Since $c^{d5}$
corresponds to neutrino masses, it must be small, implying that the
effects coming from the $d6$ operators are negligible. However, if we
look at the type-II model, we observe that we can have a relatively
large $c^{d6}$ (here we are referring to the first of the list, which
is the phenomenologically relevant one) with large Yukawa couplings
and small neutrino masses if the scale $\mu_\Delta$ is small
enough. For example, if $Y\sim \mathcal{O}(1)$ and $M\sim1$TeV, then
$\mu_\Delta\sim1$eV would give the correct order of magnitude for
neutrino masses. In this model $\mu_\Delta$ is the dimension-full
trilinear coupling among two Higgs doublets and the new scalar triplet
and it is the scale associated to lepton number violation. Since $d6$
operators are lepton number conserving, we can
suppose~\cite{Abada:2007ux} that whenever the lepton number breaking
scale $\mu$ is small and does not coincide with the scale of NP $M$,
neutrino masses will be directly proportional to $\mu$, while $c^{d6}$
will remain insensitive to this:
\be
\label{ansatz}
c^{d5} = f(Y) \frac{ \mu } {  M^2} \quad\quad\quad
c^{d6} = g(Y) \, \frac{1}{|M|^2} \, ,
\ee
where $f$ and $g$ are some functions of Yukawa couplings.  If $M$ is
low enough, effects coming from NP responsible for neutrino masses can
be observed without invoking any fine-tuning, since the $d5$ and $d6$
operators are now decoupled. As already mentioned, this is
``naturally''realized in the type-II seesaw model. As for fermionic
seesaws, particular textures of heavy Majorana masses have to be
chosen~\cite{inverseseesaw}, but it is still feasible.

Once established that the seesaw scale can be lowered without
introducing any fine-tuning to maintain neutrino masses small, the
associated effects produced by different $d6$ operators can be
analyzed. The kinetic corrections given by the $d6$ operator in the
type-I and type-III models produce a non-unitary mixing matrix in the
SM charged currents and flavor changing neutral currents (FCNC) in
the neutrino sector~\cite{BGJ,Antusch:2006vwa}. Moreover, only in the
type-III case, FCNC are present in the charged lepton sector
too~\cite{Abada:2007ux}. Among the three $d6$ operators characteristic
of the type-II model, only the first one is phenomenologically
relevant, since the others only renormalize the SM parameters: this
operator generates flavor-changing four-fermion interactions that can
be easily tested. In ref.~\cite{Abada:2007ux} a detailed analysis of EW
decays and flavor-changing processes in the three models have been
carried out and bounds on $c^{d6}$ have been derived. The general
trend is the following: for $M\sim 1$~TeV, $Y<10^{-1}$ or smaller.

Before concluding we discuss briefly some peculiarities of the
phenomenology of the type-III seesaw model. The presence of FCNC in
the charged lepton sector allows processes like $\mu\to eee$ and
flavor violating $\tau$ decays already at tree-level. This permits to
derive strong bounds on  the off-diagonal elements of the matrix
$c^{d6}_{\alpha\beta}$, stronger than the ones that can be derived
from radiative processes. However, in ref.~\cite{Abada:2008ea} the
$\mu\to e \gamma$ and $\tau\to l\gamma$ decays have been calculated
and the following relationship among branching ratios has been derived:
\be
Br(\mu \rightarrow e \gamma)=
1.3 \cdot 10^{-3} \cdot Br(\mu \rightarrow eee) 
\label{relation1C}
\ee
and similarly for $\tau$ decays.  This relationship is very
interesting because, as far as we know, this is the only model in
which radiative decays are so much suppressed with respect to
non-radiative ones. Moreover, since the current bound on
$Br(\mu \rightarrow eee)$ is $10^{-12}$ and the reach of future
experiments on $\mu\to e \gamma$ is $10^{-14}$, a positive signal from
these experiments would rule out this model as the unique source of
lepton flavor violation.

Additionally in ref.~\cite{Biggio:2008in} the anomalous magnetic
moment of leptons has been calculated and it has been shown that, even
if the scale $M$ is pushed down to the EW scale, the contribution of
the type-III seesaw is still smaller than (or comparable with) the SM
error on this observable, rendering thus impossible to measure
it. Moreover it turns out that this contribution has the ``wrong''
sign in order to explain the measured discrepancy in the muon sector.

To summarize, we have shown that the physical processes associated to
the $d6$ operators coming from different seesaw models can be used to
distinguish among them, if the scale of new physics is low enough. We
have shown that this is possible if neutrino masses suppression is due
to a small scale at which lepton number breaking is realized,
different from the scale of the new heavy particle. Finally some
details on the phenomenology of the type-III seesaw model have been
discussed.


\end{document}